\documentclass[aps,pra,preprint,superscriptaddress]{revtex4-2}
\usepackage{amsmath}
\usepackage{graphicx}
\begin{document}

% Use the \preprint command to place your local institutional report
% number in the upper righthand corner of the title page in preprint mode.
% Multiple \preprint commands are allowed.
% Use the 'preprintnumbers' class option to override journal defaults
% to display numbers if necessary
%\preprint{}

%Title of paper
\title{No circular birefringence in a chiral medium: an analysis of single-mode refraction}

% repeat the \author .. \affiliation  etc. as needed
% \email, \thanks, \homepage, \altaffiliation all apply to the current
% author. Explanatory text should go in the []'s, actual e-mail
% address or url should go in the {}'s for \email and \homepage.
% Please use the appropriate macro foreach each type of information

% \affiliation command applies to all authors since the last
% \affiliation command. The \affiliation command should follow the
% other information
% \affiliation can be followed by \email, \homepage, \thanks as well.
\author{Zhi-Juan Hu}
\email[]{huzhijuan@shnu.edu.cn}
%\homepage[]{Your web page}
%\thanks{}
%\altaffiliation{}
\affiliation{Department of Physics, Shanghai Normal University, 100 Guilin Road, 200233 Shanghai, China}

\author{Chun-Fang Li}
\email[Corresponding author: ]{cfli@shu.edu.cn}
%\homepage[]{Your web page}
%\thanks{}
%\altaffiliation{}
\affiliation{Department of Physics, Shanghai University, 99 Shangda Road, 200444 Shanghai, China}

%Collaboration name if desired (requires use of superscriptaddress
%option in \documentclass). \noaffiliation is required (may also be
%used with the \author command).
%\collaboration can be followed by \email, \homepage, \thanks as well.
%\collaboration{}
%\noaffiliation

%\date{\today}

\begin{abstract}

It is generally believed that the right-handed circularly polarized (RCP) and left-handed circularly polarized (LCP) waves in an isotropic chiral medium propagate at different velocities, known as circular birefringence. Here we show that this is not the case.
After obtaining the refraction and reflection coefficients of any elliptically polarized wave at the surface of a chiral medium, we derive the conditions for single-mode refraction.
By means of the process of single-mode refraction, we demonstrate that both the refracted RCP and the refracted LCP waves at normal incidence can be expressed as a coherent superposition of a pair of orthogonal linearly polarized waves that are rotated simultaneously. As a result, they must propagate at the same velocity as the linearly polarized waves.
A physical interpretation is also given in detail. In particular, we show that the state of polarization of any elliptically polarized wave in a chiral medium is rotated with propagation. Such a rotation amounts to the rotation of polarization bases without involving the change of the Jones vector.
The rotation of the RCP and LCP waves, as special cases of elliptically polarized waves, results in two opposite phases as if they propagated at different phase velocities with their polarization states transmitted unchanged.

\end{abstract}

% insert suggested keywords - APS authors don't need to do this
%\keywords{}

%\maketitle must follow title, authors, abstract, and keywords

\maketitle

% body of paper here - Use proper section commands
% References should be done using the \cite, \ref, and \label commands

\newpage

\section{Introduction}

A chiral medium is an optically active material \cite{Hecht}. It has the ability to rotate the plane of polarization of a linearly polarized light wave. Such an ability is associated with the absence of mirror symmetry of its molecular or crystalline configuration \cite{Barr}. A helically-coiled optical fiber also behaves as a chiral medium \cite{Papp-H, Ulri-S, Ross, Chen-R}.
Chiral objects appear in either right- or left-handed form. They provide a coupling of the magnetic field to the electric polarizability and of the electric field to the magnetic polarizability, which is described macroscopically by the constitutive relations.
A well-accepted set of constitutive relations for an isotropic and transparent chiral medium can be written as follows \cite{Geor, Bass-PE, Silv},
\begin{subequations}\label{CR}
	\begin{align}
		\mathbf{D} &= \varepsilon \mathbf{E}-g \partial{\mathbf{H}}/\partial{t}, \label{D-H} \\
		\mathbf{B} &= \mu \mathbf{H}+g \partial{\mathbf{E}}/\partial{t},         \label{B-E}
	\end{align}
\end{subequations}
where $\mathbf E$, $\mathbf H$, $\mathbf D$, and $\mathbf B$ are, as usual, the vectors of electric field, magnetic field, electric displacement, and magnetic induction, respectively, $\varepsilon$ is the permittivity, $\mu$ is the permeability, and the pseudo-scalar constant $g$ is the gyrotropic parameter.
As early as in 1825, Fresnel \cite{Hecht} proposed a simple phenomenological description of optical activity. Since a linearly polarized wave can be expressed as a coherent superposition of right-handed circularly polarized (RCP) and left-handed circularly polarized (LCP) waves, he suggested that these two circularly polarized waves propagate at different velocities. A chiral medium shows allogyric birefringence \cite{Ditc} or circular birefringence; that is to say, it possesses two indices of refraction, one for RCP wave and the other for LCP wave. Nowadays, circular birefringence has seemed to be widely accepted as true \cite{Lakh-VV, Silv, Bass-PE, Cory-R, Geor, Vale-BSV, Alex-BLY}. In particular, the difference between these two indices of refraction is even defined \cite{Barr, Kami, Ghos-F, Pfei-F, Xi-WWB, Alex-LMY, Alex-BLVY} plainly as the circular birefringence.
We will show here that this is not the case.

It is indeed true, as demonstrated recently by Ghosh and Fischer \cite{Ghos-F}, that circular double refraction \cite{Note} occurs at the planar surface of a chiral liquid though it is isotropic.
The problem, however, is that at oblique incidence, the two refracted circularly polarized modes do not propagate in the same direction. They cannot combine to yield a linearly polarized wave no matter what the incidence polarization is. In order for the refracted wave in a chiral medium to be linearly polarized, one needs to make use of normal incidence \cite{Bass-PE, Cory-R, Ghos-F}. The incident wave in this case should also be linearly polarized.
Furthermore, single-mode refraction is permissible at the surface of a chiral medium. That is to say, one of the refracted circularly polarized modes vanishes under suitable conditions.
As a matter of fact, Jaggard and Sun \cite{Jagg-S} predicted that at any incidence angle there exist two states of incidence polarization for single-mode refraction, one for each handedness of circular polarizations.
By this it is meant that given the incidence angle, changing only the incidence polarization from one state to another will convert the refracted wave from RCP to LCP or the other way around. So if the two circularly polarized waves in a chiral medium propagate at different velocities, one will be able to alter the velocity of the refracted wave by changing only the incident wave from one particular polarization state to another. This is rather puzzling.

According to Ref. \cite{Jagg-S}, the conditions for single-mode refraction require the incident wave at normal incidence to be circularly polarized.
It is known that an incident circularly polarized wave in the achiral region can be represented as a coherent superposition of two orthogonal linearly polarized waves. It is also known that an incident linearly polarized wave normal to the surface of a chiral medium gives rise to a refracted linearly polarized wave. So at normal incidence, either of the refracted circularly polarized waves under the conditions for single-mode refraction can be represented as a coherent superposition of two orthogonal linearly polarized waves.
This means that the velocities of the two circularly polarized waves in a chiral medium are the same as that of their linearly polarized components and thus cannot be different. That is to say, there is not circular birefringence at all in a chiral medium.
The key point here is that the polarization states of circularly polarized waves in a chiral medium are rotated in the same way as those of linearly polarized or any other elliptically polarized waves.
After a same propagation distance, the rotation of the polarization states of the RCP and LCP waves yields two opposite phases as if they propagated at different phase velocities with their polarization states transmitted unchanged.
The purpose of the present article is to report these new findings. The contents are arranged as follows.

In Section \ref{RRC}, we derive the refraction and reflection coefficients of any elliptically polarized wave at the surface of an isotropic chiral medium. To do this, we expand the polarization vector of the incident wave in terms of circular-polarization bases. The combination of the expansion coefficients plays the role of specifying the incidence polarization. As expected, the refraction and reflection coefficients depend not only on the incidence angle but also on the incidence polarization.
Conditions for single-mode refraction are deduced in Section \ref{SMR} from the refraction coefficients. Conditions under which the reflected wave is circularly polarized are also given. The main result of this paper is put forward in Section \ref{NCB}. It is demonstrated that whether the RCP or the LCP wave in a chiral medium can be expressed as a coherent superposition of two orthogonal linearly polarized waves. They should propagate at the velocity at which the linearly polarized waves propagate. In other words, there cannot be circular birefringence in a chiral medium.
A physical interpretation is further given in Section \ref{FDs}. Firstly, the common phase velocity of the linearly polarized and the circularly polarized waves is discussed. Secondly, it is shown that the state of polarization of any elliptically polarized wave in a chiral medium is rotated. This is expressed by the rotatory polarization vector. Such a rotation amounts to the rotation of polarization bases. The result of the rotation of the RCP and LCP waves, as special cases of elliptically polarized waves, appears mathematically as two opposite phases as if they propagated at different phase velocities with their polarization states transmitted unchanged. It is these two opposite phases that account for the circular double refraction at the surface of a chiral medium.
Section \ref{conclusions} summarizes the paper with remarks.

\section{Refraction and reflection coefficients of any elliptically polarized waves}\label{RRC}

Jaggard and Sun \cite{Jagg-S} once gave the conditions for single-mode refraction at the interface between two different chiral media where the incident wave is a coherent superposition of RCP and LCP components that propagate in different directions.
For our purpose, we will re-derive these conditions at the interface between the air of permittivity $\varepsilon_0$ and permeability $\mu_0$ and an isotropic chiral medium such as the chiral liquid considered in Ref. \cite{Ghos-F}.
A well-known fact is that single-mode refraction is permissible at the surface of a dielectric medium, which is achiral. The refracted characteristic wave in that case is linearly polarized. In particular, the incident and refracted waves under the conditions for single-mode refraction have the same polarization. They are either $s$-polarized or $p$-polarized.
Probably influenced by that fact, some authors \cite{Ghos-F} thought that at the surface of a chiral medium, the incident and refracted waves under the conditions for single-mode refraction also have the same polarization. They are either RCP or LCP. However this is incorrect.
To avoid such a misunderstanding, it is better to have the refraction and reflection coefficients of any elliptically polarized wave.

\subsection{Expressions for incident and reflected waves}

\begin{figure}[t]
	\centerline{\includegraphics[width=7.5cm]{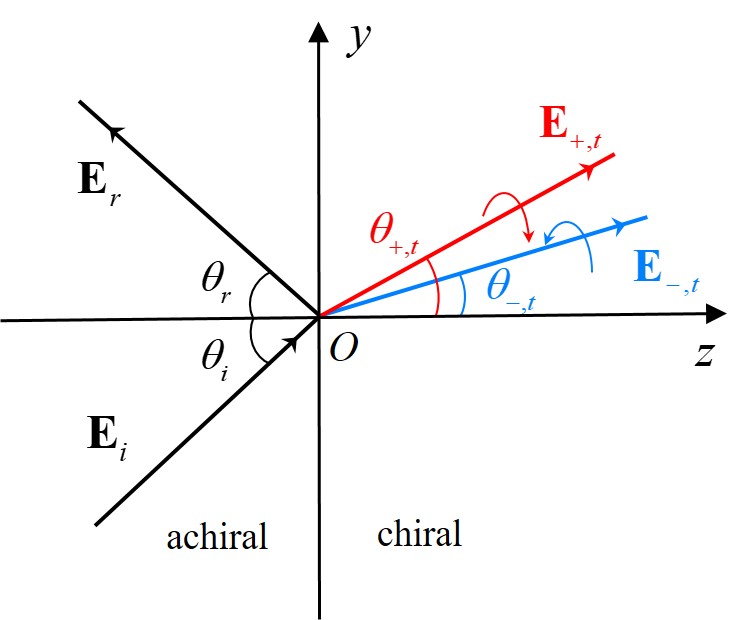}}
	\caption{\label{CDR} Illustration for the circular double refraction at the surface of an isotropic chiral medium.}
\end{figure}
As is shown in Fig. \ref{CDR}, a plane wave is incident from the air upon the surface of the chiral medium at the incidence angle $\theta_i$.
Considering that the refracted characteristic modes are circularly polarized, we expand the incident wave in terms of RCP and LCP components rather than in terms of $s$- and $p$-polarized components as is usually done in the literature \cite{Bass-PE, Geor, Silv, Lakh-VV}.
Letting $\bar x$, $\bar y$, and $\bar z$ be the unit vectors along the corresponding Cartesian axes, the electric field of the incident wave can be expressed as follows,
\begin{equation}\label{Ei}
	\mathbf{E}_i (\mathbf{x})=\bigg[\frac{c_{+}}{\sqrt 2}(\mathbf{s}_i +i \mathbf{p}_i)
                              +\frac{c_{-}}{\sqrt 2}(\mathbf{s}_i -i \mathbf{p}_i) \bigg]
	                           E_i \exp(i \mathbf{k}_i \cdot \mathbf{x}),
\end{equation}
where the time-dependence $\exp(-i\omega t)$ is assumed,
$\mathbf{k}_i =k_0 (\bar{y} \sin \theta_i +\bar{z} \cos \theta_i)$ is the wavevector,
$k_0 =(\varepsilon_0 \mu_0)^{1/2} \omega$, $\mathbf{s}_i =\bar{x}$,
$\mathbf{p}_i =\frac{\mathbf{k}_i}{k_0} \times \mathbf{s}_i =\bar{y} \cos \theta_i -\bar{z} \sin \theta_i$,
the expansion coefficients $c_{+}$ and $c_{-}$ satisfy
\begin{equation}\label{normalization}
	|c_{+}|^2 +|c_{-}|^2 =1,
\end{equation}
and $E_i$ is the amplitude. The unit vectors
$\frac{1}{\sqrt 2} (\mathbf{s}_i +i \mathbf{p}_i)$ and $\frac{1}{\sqrt 2} (\mathbf{s}_i -i \mathbf{p}_i)$
in expression (\ref{Ei}) are the circular-polarization bases. They stand for the RCP and LCP states, respectively. So the combination of $c_{+}$ and $c_{-}$ plays the role of specifying the incidence polarization, giving the polarization ellipticity
$\sigma =|c_{+}|^2-|c_{-}|^2$. When $c_{-}=0$, it is RCP. When $c_{+}=0$, it is LCP.
According to Maxwell's equation
$\nabla \times \mathbf{E}= -\partial \mathbf{B}/\partial t$
and the constitutive relation
$\mathbf{B}= \mu_0 \mathbf{H}$,
the magnetic field of the incident wave takes the form
\begin{equation*}%\label{Hi}
	\mathbf{H}_i (\mathbf{x})=\frac{1}{i \eta_0}
	                          \bigg[ \frac{c_{+}}{\sqrt 2}(\mathbf{s}_i +i \mathbf{p}_i) 
	                                -\frac{c_{-}}{\sqrt 2}(\mathbf{s}_i -i \mathbf{p}_i) \bigg]
	                          E_i \exp(i \mathbf{k}_i \cdot \mathbf{x}),
\end{equation*}
where $\eta_0 =\sqrt{\mu_0/\varepsilon_0}$.
Similarly, letting be $\theta_r$ the reflection angle, the electric field of the reflected wave can be expressed as
\begin{equation}\label{Er}
	\mathbf{E}_r (\mathbf{x})=\bigg[ \frac{E_{+,r}}{\sqrt 2} (\mathbf{s}_r +i \mathbf{p}_r)
                             +\frac{E_{-,r}}{\sqrt 2} (\mathbf{s}_r -i \mathbf{p}_r) \bigg]
	                          \exp(i \mathbf{k}_r \cdot \mathbf{x}),
\end{equation}
where
$\mathbf{k}_r =k_0 (\bar{y} \sin \theta_r -\bar{z} \cos \theta_r)$,
$\mathbf{s}_r =\bar{x}$,
$\mathbf{p}_r =\frac{\mathbf{k}_r}{k_0} \times \mathbf{s}_r =-\bar{y} \cos \theta_r -\bar{z} \sin \theta_r$,
and $E_{+,r}$ and $E_{-,r}$ are the amplitudes of the RCP and LCP components, respectively.
Here the unit vectors $\frac{1}{\sqrt 2}(\mathbf{s}_r +i \mathbf{p}_r)$ and $\frac{1}{\sqrt 2}(\mathbf{s}_r -i \mathbf{p}_r)$ stand for the RCP and LCP states in the reflected wave, respectively.
Accordingly, the magnetic field of the reflected wave assumes
\begin{equation*}%\label{Hr}
	\mathbf{H}_r (\mathbf{x})=\frac{1}{i \eta_0}
	                          \bigg[ \frac{E_{+,r}}{\sqrt 2}(\mathbf{s}_r +i \mathbf{p}_r) 
	                                -\frac{E_{-,r}}{\sqrt 2}(\mathbf{s}_r -i \mathbf{p}_r) \bigg]
	                          \exp(i \mathbf{k}_r \cdot \mathbf{x}).
\end{equation*}

It is pointed out that expression (\ref{Er}) for the reflected wave is not parallel to expression (\ref{Ei}) for the incident wave. This is because the incident wave is supposed to be known so that both the amplitude $E_i$ and the expansion coefficients $c_{+}$ and $c_{-}$ are given; whereas the reflected wave needs to be determined by use of the boundary conditions.
We will see below that what we obtain directly from the boundary conditions are $E_{+,r}$ and $E_{-,r}$. If defined as
$\frac{E_{+,r}}{E_i}$ and $\frac{E_{-,r}}{E_i}$, the reflection coefficients will depend on the incidence polarization specified by $c_{+}$ and $c_{-}$.

\subsection{Expressions for refracted wave in the chiral medium}

As was experimentally observed \cite{Ghos-F}, the refracted wave consists of RCP and LCP components that propagate in different directions. According to constitutive relations (\ref{CR}), the ``wave numbers'' of the RCP and LCP components are given by \cite{Silv,Geor}
\begin{equation}\label{k+andk-}
	\begin{split}
		k_{+,t} &= k-\tau,\\
		k_{-,t} &= k+\tau, 
	\end{split}
\end{equation}
respectively, where $k=(\varepsilon \mu)^{1/2} \omega$ and $\tau=-g \omega^2$.
Letting be $\theta_{+,t}$ the refraction angle of the RCP component, its electric field can be expressed as
\begin{equation*}
	\mathbf{E}_{+,t} (\mathbf{x})=\frac{E_{+,t}}{\sqrt 2} (\mathbf{s}_{+,t}+i\mathbf{p}_{+,t})
	                              \exp(i\mathbf{k}_{+,t} \cdot \mathbf{x}),
\end{equation*}
where $E_{+,t}$ is the amplitude,
$\mathbf{k}_{+,t}=k_{+,t} (\bar{y} \sin \theta_{+,t} +\bar{z} \cos \theta_{+,t})$ is the wavevector,
$\mathbf{s}_{+,t} =\bar{x}$, and
$\mathbf{p}_{+,t} =\frac{\mathbf{k}_{+,t}}{k_{+,t}} \times \mathbf{s}_{+,t}
                  =\bar{y} \cos \theta_{+,t} -\bar{z} \sin \theta_{+,t}$.
The magnetic field of the RCP component is found, upon making use of Maxwell's equation
$\nabla \times \mathbf{E}= -\partial \mathbf{B}/\partial t$ and constitutive relation (\ref{B-E}), to be
$\mathbf{H}_{+,t} (\mathbf{x}) =-\frac{ik}{\mu \omega} \mathbf{E}_{+,t} (\mathbf{x})$
and is given by
\begin{equation*}
	\mathbf{H}_{+,t} (\mathbf{x})=\frac{E_{+,t}}{i \sqrt{2} \eta} (\mathbf{s}_{+,t}+i\mathbf{p}_{+,t})
	                              \exp(i\mathbf{k}_{+,t} \cdot \mathbf{x}),
\end{equation*}
where $\eta=\sqrt{\mu/\varepsilon}$.
Similarly, letting be $\theta_{-,t}$ the refraction angle of the LCP component, its electric field can be expressed as
\begin{equation*}
	\mathbf{E}_{-,t} (\mathbf{x})=\frac{E_{-,t}}{\sqrt 2} (\mathbf{s}_{-,t}-i\mathbf{p}_{-,t})
	                              \exp(i\mathbf{k}_{-,t} \cdot \mathbf{x}),
\end{equation*}
where $E_{-,t}$ is the amplitude,
$\mathbf{k}_{-,t}=k_{-,t} (\bar{y} \sin \theta_{-,t} +\bar{z} \cos \theta_{-,t})$,
$\mathbf{s}_{-,t} =\bar{x}$, and
$\mathbf{p}_{-,t} =\frac{\mathbf{k}_{-,t}}{k_{-,t}} \times \mathbf{s}_{-,t}
                  =\bar{y} \cos \theta_{-,t} -\bar{z} \sin \theta_{-,t}$.
Correspondingly, the magnetic field of the LCP component is found to be
$\mathbf{H}_{-,t} (\mathbf{x}) =\frac{ik}{\mu \omega} \mathbf{E}_{-,t} (\mathbf{x})$
and is given by
\begin{equation*}
	\mathbf{H}_{-,t} (\mathbf{x})=\frac{iE_{-,t}}{\sqrt{2} \eta} (\mathbf{s}_{-,t}-i\mathbf{p}_{-,t})
                                  \exp(i\mathbf{k}_{-,t} \cdot \mathbf{x}).
\end{equation*}
Consequently, the total electric field of the refracted wave,
$\mathbf{E}_t =\mathbf{E}_{+,t}+\mathbf{E}_{-,t}$,
reads
\begin{equation}\label{Et}
	\mathbf{E}_t = \frac{E_{+,t}}{\sqrt 2}(\mathbf{s}_{+,t}+i\mathbf{p}_{+,t}) 
	               \exp(i\mathbf{k}_{+,t} \cdot \mathbf{x})
	              +\frac{E_{-,t}}{\sqrt 2}(\mathbf{s}_{-,t}-i\mathbf{p}_{-,t}) 
	               \exp(i\mathbf{k}_{-,t} \cdot \mathbf{x}),
\end{equation}
and the total magnetic field,
$\mathbf{H}_t =\mathbf{H}_{+,t}+\mathbf{H}_{-,t}$,
reads
\begin{equation*}
	\mathbf{H}_t =\frac{1}{i \eta}
	              \bigg[ \frac{E_{+,t}}{\sqrt 2}(\mathbf{s}_{+,t}+i\mathbf{p}_{+,t}) 
	                     \exp(i\mathbf{k}_{+,t} \cdot \mathbf{x})
	                    -\frac{E_{-,t}}{\sqrt 2}(\mathbf{s}_{-,t}-i\mathbf{p}_{-,t}) 
	                     \exp(i\mathbf{k}_{-,t} \cdot \mathbf{x}) \bigg].
\end{equation*}

\subsection{Refraction and reflection coefficients}

Having written out the electric and magnetic fields of the incident, reflected, and refracted waves, we are in a position to make use of the boundary conditions to determine the reflection angle $\theta_r$ as well as the refraction angles $\theta_{+,t}$ and $\theta_{-,t}$ and to find the reflection and refraction coefficients.
The phase-matching condition at the interface between the air and the chiral medium leads to Snell's law
\begin{equation}\label{RL}
	\sin \theta_r =\sin \theta_i
\end{equation}
for the reflection angle $\theta_r$ as well as Snell's laws
\begin{equation}\label{SL}
	k_{+,t} \sin \theta_{+,t} =k_{-,t} \sin \theta_{-,t} =k_0 \sin \theta_i
\end{equation}
for the refraction angles $\theta_{+,t}$ and $\theta_{-,t}$. After these equations are taken into account,
the continuity of the $x$-component of the electric field at the interface yields
\begin{equation}\label{E-y}
	(c_{+} +c_{-}) E_i +E_{+,r} +E_{-,r} =E_{+,t} +E_{-,t}
\end{equation}
and the continuity of the $y$-component of the electric field at the interface yields
\begin{equation}\label{E-z}
	(c_{+} -c_{-}) E_i -E_{+,r} +E_{-,r} =\frac{\cos \theta_{+,t}}{\cos \theta_i} E_{+,t}
	                                     -\frac{\cos \theta_{-,t}}{\cos \theta_i} E_{-,t}.
\end{equation}
Similarly, after Eqs. (\ref{RL}) and (\ref{SL}) are taken into account, the continuity of the $x$-component of the magnetic field at the interface means
\begin{equation}\label{H-y}
	\frac{1}{\eta_0} [(c_{+} -c_{-}) E_i +E_{+,r} -E_{-,r}] =\frac{1}{\eta} (E_{+,t} -E_{-,t})
\end{equation}
and the continuity of the $y$-component of the magnetic field at the interface means
\begin{equation}\label{H-z}
	\frac{1}{\eta_0} [(c_{+} +c_{-}) E_i -E_{+,r} -E_{-,r}]
   =\frac{1}{\eta} \Big(\frac{\cos \theta_{+,t}}{\cos \theta_i} E_{+,t}
                       +\frac{\cos \theta_{-,t}}{\cos \theta_i} E_{-,t}\Big).
\end{equation}

From Eqs. (\ref{E-y})-(\ref{H-z}) one readily finds the refraction coefficients for the two characteristic modes, which are defined as
$\frac{E_{+,t}}{E_i} \equiv T_{+}$ and $\frac{E_{-,t}}{E_i} \equiv T_{-}$
and are given by
\begin{subequations}\label{TCs}
	\begin{align}
		T_{+} &= 2 \cos \theta_i
		         \frac{(n+1)(\cos \theta_{-,t} +\cos \theta_i) c_{+}
			          -(n-1)(\cos \theta_{-,t} -\cos \theta_i) c_{-}}
		              {2n(\cos^2 \theta_i +\cos \theta_{+,t} \cos \theta_{-,t})
		              +(n^2+1)(\cos \theta_{+,t} +\cos \theta_{-,t}) \cos \theta_i}, \label{TC1} \\
		T_{-} &= 2 \cos \theta_i
		         \frac{(n+1)(\cos \theta_{+,t} +\cos \theta_i) c_{-}
        		      -(n-1)(\cos \theta_{+,t} -\cos \theta_i) c_{+}}
		              {2n(\cos^2 \theta_i +\cos \theta_{+,t} \cos \theta_{-,t})
			          +(n^2+1)(\cos \theta_{+,t} +\cos \theta_{-,t}) \cos \theta_i}, \label{TC2}
	\end{align}
\end{subequations}
where $n=\eta_0/\eta$. So defined refraction coefficients depend not only on the incidence angle $\theta_i$ but also on the incidence polarization specified by $c_{+}$ and $c_{-}$. It is worth noting that such a definition gets rid of the so-called cross-polarized refraction coefficients \cite{Jagg-S}.
From Eqs. (\ref{E-y})-(\ref{H-z}) one can also find the reflection coefficients for the two circularly polarized components, which are defined as
$\frac{E_{+,r}}{E_i} \equiv R_{+}$ and $\frac{E_{-,r}}{E_i} \equiv R_{-}$
and are given by
\begin{subequations}\label{RCs}
	\begin{align}
		R_{+} &= \frac{2n(\cos \theta_i +\cos \theta_{-,t}) (\cos \theta_i -\cos \theta_{+,t}) c_{+}
			          -(n^2-1) \cos \theta_i (\cos \theta_{+,t} +\cos \theta_{-,t}) c_{-}}
		              {2n(\cos^2 \theta_i +\cos \theta_{+,t} \cos \theta_{-,t})
			          +(n^2+1)(\cos \theta_{+,t} +\cos \theta_{-,t}) \cos \theta_i}, \label{RC1} \\
		R_{-} &= \frac{2n(\cos \theta_i +\cos \theta_{+,t}) (\cos \theta_i -\cos \theta_{-,t}) c_{-}
			          -(n^2-1) \cos \theta_i (\cos \theta_{+,t} +\cos \theta_{-,t}) c_{+}}
		              {2n(\cos^2 \theta_i +\cos \theta_{+,t} \cos \theta_{-,t})
			          +(n^2+1)(\cos \theta_{+,t} +\cos \theta_{-,t}) \cos \theta_i}. \label{RC2}
	\end{align}
\end{subequations}
Akin to the refraction coefficients, so defined reflection coefficients depend not only on the incidence angle but also on the incidence polarization.

\section{Conditions for single-mode refraction}\label{SMR}

In general, double refraction occurs at the surface of the chiral medium. The refracted wave consists of RCP and LCP components that propagate in different directions. But when the incidence angle and the state of incidence polarization satisfy certain conditions, only one mode, either RCP or LCP, exists in the refracted wave. In fact, expression (\ref{TC2}) tells that only the RCP mode will be refracted if
\begin{equation}\label{SMT1}
	\frac{c_{-}}{c_{+}}=\frac{n-1}{n+1}
	                    \frac{\cos \theta_{+,t}-\cos \theta_i}{\cos \theta_{+,t}+\cos \theta_i}.
\end{equation}
On the other hand, expression (\ref{TC1}) tells that only the LCP mode will be refracted if
\begin{equation}\label{SMT2}
	\frac{c_{+}}{c_{-}}=\frac{n-1}{n+1}
	\frac{\cos \theta_{-,t}-\cos \theta_i}{\cos \theta_{-,t}+\cos \theta_i}.
\end{equation}
Clearly, at any incidence angle there exist two states of incidence polarization for single-mode refraction, one for each handedness of circular polarizations.
Since usually $n \ne 1$, condition (\ref{SMT1}) does not mean $c_{-} =0$ and condition (\ref{SMT2}) does not mean $c_{+} =0$ unless the incidence angle is equal to zero. This shows that at oblique incidence the incidence polarization for single-mode refraction is not circular.
Frankly speaking, one has $n>1$ for conventional chiral media. In this case, condition (\ref{SMT1}) means $|c_{-}| <|c_{+}|$, the incident wave being right-handed elliptically polarized. In contrast, condition (\ref{SMT2}) means $|c_{+}| <|c_{-}|$, the incident wave being left-handed elliptically polarized.

Of course, the reflected wave in the air is in general elliptically polarized. But Eqs. (\ref{RCs}) show that there exist conditions under which the reflected wave is circularly polarized. Eq. (\ref{RC2}) means that the reflected wave will be RCP if
\begin{equation*}
	\frac{c_{+}}{c_{-}}=\frac{2n}{n^2 -1}
	\frac{(\cos \theta_i +\cos \theta_{+,t})(\cos \theta_i -\cos \theta_{-,t})}
	{(\cos \theta_{+,t} +\cos \theta_{-,t}) \cos \theta_i}.
\end{equation*}
Similarly, Eq. (\ref{RC1}) means that the reflected wave will be LCP if
\begin{equation*}
	\frac{c_{-}}{c_{+}}=\frac{2n}{n^2 -1}
	\frac{(\cos \theta_i -\cos \theta_{+,t})(\cos \theta_i +\cos \theta_{-,t})}
	{(\cos \theta_{+,t} +\cos \theta_{-,t}) \cos \theta_i}.
\end{equation*}

\section{There is not circular birefringence}\label{NCB}

According to conditions (\ref{SMT1}) and (\ref{SMT2}) for single-mode refraction, at any incidence angle there exist two states of incidence polarization for the refracted wave to be circularly polarized. Importantly, the incidence polarization at normal incidence is required to be circular, too. This provides a possibility for us to logically analyze the propagation velocities of the RCP and LCP waves in the chiral medium.

Since the incident circularly polarized wave can be expressed as a coherent superposition of two orthogonal linearly polarized waves, the refracted circularly polarized wave could also be expressed as a coherent superposition of two orthogonal linearly polarized waves that are rotated in the same way. To show this, let us first find out such two refracted linearly polarized waves.
When $\theta_i =0$, expression (\ref{Ei}) for the electric field of the incident wave reads
\begin{equation}\label{Ei-NI}
	\mathbf{E}_i (\mathbf{x})=\frac{1}{\sqrt 2}
	                         [\bar{x} (c_{+}+c_{-})	+i \bar{y} (c_{+}-c_{-})] E_i \exp(ik_0 z).
\end{equation}
Since the angles of refraction at normal incidence are equal to zero, $\theta_{+,t}=\theta_{-,t}=0$, the refraction coefficients (\ref{TCs}) become
\begin{eqnarray*}
% \nonumber to remove numbering (before each equation)
  T_{+} &=& T c_{+}, \\
  T_{-} &=& T c_{-},
\end{eqnarray*}
where
$T=\frac{2}{n+1}$.
As a result, expression (\ref{Et}) for the electric field of the refracted wave assumes
\begin{equation}\label{Et-NI}
	\mathbf{E}_t (\mathbf{x})
   =\frac{1}{\sqrt 2}[(\bar{x} \cos \tau z +\bar{y} \sin \tau z) (c_{+}+c_{-})
                       +i(\bar{y} \cos \tau z -\bar{x} \sin \tau z) (c_{+}-c_{-})] T E_i \exp(ikz),
\end{equation}
by virtue of Eqs. (\ref{k+andk-}). A comparison with expression (\ref{Ei-NI}) shows that for this wave to be linearly polarized, the incident wave also needs to be linearly polarized.
In particular, when $c_{+}=c_{-}=\frac{\sqrt 2}{2}$, expression (\ref{Ei-NI}) reduces to
\begin{equation}\label{E1i}
	\mathbf{E}_{1,i}=\bar{x} E_i \exp(ik_0 z);
\end{equation}
and expression (\ref{Et-NI}) reduces to
\begin{equation}\label{E1t}
	\mathbf{E}_{1,t}=(\bar{x} \cos \tau z +\bar{y} \sin \tau z) T E_i \exp(ikz).
\end{equation}
In addition, when $c_{+}=-c_{-}=-i\frac{\sqrt 2}{2}$, expression (\ref{Ei-NI}) reduces to
\begin{equation}\label{E2i}
	\mathbf{E}_{2,i}=\bar{y} E_i \exp(ik_0 z);
\end{equation}
and expression (\ref{Et-NI}) reduces to
\begin{equation}\label{E2t}
	\mathbf{E}_{2,t}=(\bar{y} \cos \tau z -\bar{x} \sin \tau z) T E_i \exp(ikz).
\end{equation}
It is not difficult to see that the refracted waves (\ref{E1t}) and (\ref{E2t}), which are rotated with propagation in the same way, are orthogonal to each other at the same propagation distance $z$. This is understandable, because their corresponding incident waves (\ref{E1i}) and (\ref{E2i}) are orthogonal to each other.

To see further how these two linearly polarized waves can be utilized to express a circularly polarized wave in the chiral medium, we resort to the familiar fact to expand them in terms of the RCP and LCP waves.
When $c_{+}=1$ and $c_{-}=0$, the refracted wave (\ref{Et-NI}) is RCP,
\begin{equation}\label{E+t_no}
	\mathbf{E}_{+,t}=\frac{1}{\sqrt 2} (\bar{x}+i\bar{y}) T E_i \exp(ik_{+,t} z).
\end{equation}
And when $c_{+}=0$ and $c_{-}=1$, the refracted wave (\ref{Et-NI}) is LCP,
\begin{equation}\label{E-t_no}
	\mathbf{E}_{-,t}=\frac{1}{\sqrt 2}(\bar{x}-i \bar{y}) T E_i \exp(i k_{-,t}z).
\end{equation}
As expected \cite{Hecht}, the linearly polarized wave $\mathbf{E}_{1,t}$ can be expressed as a coherent superposition of the circularly polarized waves $\mathbf{E}_{+,t}$ and $\mathbf{E}_{-,t}$ of the form,
\begin{equation}\label{E1t-CS}
	\mathbf{E}_{1,t}=\frac{1}{\sqrt 2}(\mathbf{E}_{+,t} +\mathbf{E}_{-,t}).
\end{equation}
Similarly, the linearly polarized wave $\mathbf{E}_{2,t}$ can also be expressed as a coherent superposition of $\mathbf{E}_{+,t}$ and $\mathbf{E}_{-,t}$,
\begin{equation}\label{E2t-CS}
	\mathbf{E}_{2,t}=-\frac{i}{\sqrt 2}(\mathbf{E}_{+,t} -\mathbf{E}_{-,t}).
\end{equation}
From Eqs. (\ref{E1t-CS}) and (\ref{E2t-CS}) it follows that the RCP wave $\mathbf{E}_{+,t}$ can be expressed as a coherent superposition of the linearly polarized waves $\mathbf{E}_{1,t}$ and $\mathbf{E}_{2,t}$ as follows,
\begin{equation}\label{E+t_CS}
	\mathbf{E}_{+,t}=\frac{1}{\sqrt 2}(\mathbf{E}_{1,t}+i \mathbf{E}_{2,t}).
\end{equation}
In addition, the LCP wave $\mathbf{E}_{-,t}$ can also be expressed as a coherent superposition of the same two linearly polarized waves of the form,
\begin{equation}\label{E-t_CS}
	\mathbf{E}_{-,t}=\frac{1}{\sqrt 2}(\mathbf{E}_{1,t}-i \mathbf{E}_{2,t}).
\end{equation}
In a word, whether the RCP wave $\mathbf{E}_{+,t}$ or the LCP wave $\mathbf{E}_{-,t}$ can be expressed as a coherent superposition of the orthogonal linearly polarized waves $\mathbf{E}_{1,t}$ and $\mathbf{E}_{2,t}$.

We are now ready to give our main result that the two refracted characteristic modes in the chiral medium do not propagate at different velocities.
Rotated in the same way, the linearly polarized waves $\mathbf{E}_{1,t}$ and $\mathbf{E}_{2,t}$ in the chiral medium must propagate at the same velocity if they have definite propagation velocities.
As a result of Eqs. (\ref{E+t_CS}) and (\ref{E-t_CS}), the two circularly polarized waves $\mathbf{E}_{+,t}$ and $\mathbf{E}_{-,t}$ should propagate at the velocity at which the linearly polarized waves propagate. That is to say, they should propagate at the same velocity.
No circular birefringence exists in the chiral medium.
What should be pointed out is that even though expression (\ref{E+t_no}) correctly describes the RCP wave in the chiral medium, it does not mean that the RCP wave is a coherent superposition of the following linearly polarized waves,
\begin{equation}\label{PNAW1}
    \begin{split}
	    \mathbf{E}^{+}_{x,t} &= \bar{x} T E_i \exp(ik_{+,t} z),\\
        \mathbf{E}^{+}_{y,t} &= \bar{y} T E_i \exp(ik_{+,t} z).
	\end{split}
\end{equation}
This is because there does not exist in the chiral medium such a linearly polarized wave the polarization plane of which does not change. Similarly, expression (\ref{E-t_no}) does not mean that the LCP wave is a coherent superposition of the following linearly polarized waves,
\begin{equation}\label{PNAW2}
    \begin{split}
	    \mathbf{E}^{-}_{x,t} &= \bar{x} T E_i \exp(ik_{-,t} z),\\
	    \mathbf{E}^{-}_{y,t} &= \bar{y} T E_i \exp(ik_{-,t} z).
	\end{split}
\end{equation}
It is also noted that if the RCP and LCP waves $\mathbf{E}_{+,t}$ and $\mathbf{E}_{-,t}$ were considered to propagate at different velocities, one could not deduce from Eqs. (\ref{E1t-CS}) and (\ref{E2t-CS}) that the two linearly polarized waves $\mathbf{E}_{1,t}$ and $\mathbf{E}_{2,t}$ have the same propagation velocity.
A further interpretation of this result is given below.

\section{Further discussions}\label{FDs}

Let us first look at the propagation velocity of circularly polarized waves in the chiral medium. To this end, we rewrite linearly polarized waves (\ref{E1t}) and (\ref{E2t}) explicitly as
\begin{subequations}\label{LPTWs}
	\begin{align}
		\mathbf{E}_{1,t} &= \bar{x}'(z) T E_i \exp(ikz), \label{LPTW1} \\
		\mathbf{E}_{2,t} &= \bar{y}'(z) T E_i \exp(ikz), \label{LPTW2}
	\end{align}
\end{subequations}
where
\begin{equation}\label{uplusv}
	\begin{split}
		\bar{x}'(z) &= \exp[-i(\mathbf{\Sigma} \cdot \bar{z}) \tau z] \bar{x},\\
		\bar{y}'(z) &= \exp[-i(\mathbf{\Sigma} \cdot \bar{z}) \tau z] \bar{y}, 
	\end{split}
\end{equation}
are the polarization vectors depending on the propagation distance $z$ and $(\Sigma_k)_{ij} =-i \epsilon_{ijk}$ with $\epsilon_{ijk}$ the Levi-Civit\'{a} pseudotensor.
The rotation of the polarization planes of linearly polarized waves (\ref{LPTWs}) is properly represented by expressions (\ref{uplusv}), the rotatory polarization vectors. That is to say, the rotation operator $\exp[-i(\mathbf{\Sigma} \cdot \bar{z}) \tau z]$ in (\ref{uplusv}) properly conveys the role of the gyrotropic parameter $g$ in constitutive relations (\ref{CR}).
This indicates that only the common phase factor $\exp(ikz)$ in (\ref{LPTWs}) is the propagation factor. It determines the phase velocity to be $\frac{\omega}{k}$.
Substituting Eqs. (\ref{LPTWs}) into Eqs. (\ref{E+t_CS}) and (\ref{E-t_CS}), we have
\begin{subequations}\label{CPTWs}
	\begin{align}
		\mathbf{E}_{+,t} &= \mathbf{r}(z) T E_i \exp(ikz)  \label{CPTW1} \\
		\mathbf{E}_{-,t} &= \mathbf{l}(z) T E_i \exp(ikz), \label{CPTW2}
	\end{align}
\end{subequations}
where
\begin{equation}\label{rplusl}
	\begin{split}
		\mathbf{r}(z) &=\frac{1}{\sqrt 2} \exp[-i(\mathbf{\Sigma} \cdot \bar{z})\tau z]
		                (\bar{x}+i\bar{y}), \\
		\mathbf{l}(z) &=\frac{1}{\sqrt 2} \exp[-i(\mathbf{\Sigma} \cdot \bar{z})\tau z]
		                (\bar{x}-i\bar{y}).
	\end{split}
\end{equation}
A comparison of Eqs. (\ref{rplusl}) with (\ref{uplusv}) shows that the polarization vectors of circularly polarized waves (\ref{CPTWs}) are rotated in the same way as those of linearly polarized waves (\ref{LPTWs}).
This means that the same as the states of polarization of linearly polarized waves are rotated, the states of polarization of circularly polarized waves are also rotated. Such a rotation is also represented by their rotatory polarization vectors.
That is, the factors $\mathbf{r}(z)$ and $\mathbf{l}(z)$ express the rotation of the polarization states of RCP and LCP waves, respectively.
Only the common phase factor $\exp(ikz)$ in (\ref{CPTWs}) is the propagation factor. It determines the phase velocity to be $\frac{\omega}{k}$, the same as the phase velocity of linearly polarized waves (\ref{LPTWs}).
The quantity $k=(\varepsilon \mu)^{1/2} \omega$ for the chiral medium is not ``merely a shorthand notation'' \cite{Lakh-VV90} at all. It is the wave number.

Secondly, since $\bar{x}+i\bar{y}$ and $\bar{x}-i\bar{y}$ are eigen vectors of operator $\mathbf{\Sigma} \cdot \bar{z}$ with eigen values $+1$ and $-1$, respectively, expressions (\ref{rplusl}) can be rewritten as
\begin{equation}\label{rplusl2}
	\begin{split}
		\mathbf{r}(z) &=\frac{1}{\sqrt 2} (\bar{x}+i\bar{y}) \exp(-i\tau z), \\
		\mathbf{l}(z) &=\frac{1}{\sqrt 2} (\bar{x}-i\bar{y}) \exp( i\tau z),
	\end{split}
\end{equation}
meaning that the rotation of the polarization states of RCP and LCP waves in the chiral medium gives rise to phases of equal magnitude but opposite sign.
In view of this, the circular double refraction indicated by the difference between $k_{+,t}$ and $k_{-,t}$ in (\ref{k+andk-}) just reflects these two opposite phases.
It is unreasonable to interpret the RCP and LCP waves, expressed respectively by (\ref{E+t_no}) and (\ref{E-t_no}), as propagating at different phase velocities.

Then how do we understand the rotation of circularly polarized waves in the chiral medium? As discussed above, the linearly polarized waves in (\ref{LPTWs}) are orthogonal to each other at the same propagation distance. They can serve as base modes to expand the refracted wave (\ref{Et-NI}) in the following way,
\begin{equation}\label{Et-LPBs}
	\mathbf{E}_t =\alpha_1 \mathbf{E}_{1,t} +\alpha_2 \mathbf{E}_{2,t}
	             =\mathbf{a}_t (z) T E_i \exp(ikz),
\end{equation}
where
\begin{equation}\label{a-LPBs}
	\mathbf{a}_t (z)=\alpha_1 \bar{x}'(z) +\alpha_2 \bar{y}'(z)
\end{equation}
is the polarization vector, and
$\alpha_1 =\frac{1}{\sqrt 2} (c_{+}+c_{-})$
and
$\alpha_2 =\frac{i}{\sqrt 2} (c_{+}-c_{-})$
obey
$|\alpha_1|^2 +|\alpha_2|^2 =1$ by virtue of Eq. (\ref{normalization}).
As a matter of fact, it is the unit vectors $\bar{x}'$ and $\bar{y}'$ that are orthogonal to each other at the same propagation distance as can be seen from Eqs. (\ref{uplusv}).
They serve as polarization bases to expand $\mathbf{a}_t (z)$. The expansion coefficients $\alpha_1$ and $\alpha_2$ therefore form the well-known Jones vector
$\alpha =\bigg(\begin{array}{c}
  	             \alpha_1 \\ \alpha_2
               \end{array}
         \bigg)$.
It is important to note that the polarization bases mean any two orthogonal polarization vectors that are physically permissible. Any linear combination of them such as (\ref{a-LPBs}) describes a possible state of polarization.
With this in mind, the constant vectors $\bar x$ and $\bar y$ in expressions (\ref{PNAW1}) or (\ref{PNAW2}), though they are orthogonal to each other, cannot be regarded as polarization bases for the plane wave in the chiral medium.
Because the polarization bases $\bar{x}'$ and $\bar{y}'$ are rotated with propagation in the same way, the polarization state of refracted wave (\ref{Et-LPBs}) is also rotated with propagation no matter what the Jones vector is. When
$\alpha= \frac{1}{\sqrt 2} \bigg(\begin{array}{c}
	                                 1 \\ i
                                 \end{array}
                           \bigg)$
or
$\alpha= \frac{1}{\sqrt 2} \bigg(\begin{array}{c}
	                                 1 \\ -i
                                 \end{array}
                           \bigg)$,
Eq. (\ref{Et-LPBs}) goes back to circularly polarized wave (\ref{CPTW1}) or (\ref{CPTW2}). That is, the circularly polarized waves in (\ref{CPTWs}) are just special cases of elliptically polarized wave (\ref{Et-LPBs}). 
It is incorrect to say \cite{Ditc} that circularly polarized waves are transmitted unchanged in the chiral medium.
The optical activity of helically-coiled optical fibers \cite{Papp-H, Ulri-S, Ross, Chen-R} can be discussed in a similar manner with the result that the rotation of RCP and LCP waves yields opposite phases \cite{Chia-W, Akir-C, Berr}.

By the way, the two circularly polarized waves in (\ref{CPTWs}) are also orthogonal to each other. They can serve as base modes to expand the refracted wave (\ref{Et-NI}) in the form,
\begin{equation}\label{Et-CPBs}
	\mathbf{E}_t =c_{+} \mathbf{E}_{+,t} +c_{-} \mathbf{E}_{-,t}
	             =\mathbf{a}_t (z) T E_i \exp(ikz),
\end{equation}
where
\begin{equation}\label{a-CPBs}
	\mathbf{a}_t (z)=c_{+} \mathbf{r}(z) +c_{-} \mathbf{l}(z).
\end{equation}
Here the unit vectors $\mathbf r$ and $\mathbf l$, which are given in (\ref{rplusl}), also serve as polarization bases to expand the polarization vector $\mathbf{a}_t (z)$ because they are orthogonal to each other at the same propagation distance.
When $c_{+}=c_{-}=\frac{\sqrt 2}{2}$, Eq. (\ref{Et-CPBs}) goes back to linearly polarized wave (\ref{LPTW1}). Whereas when $c_{+}=-c_{-}=-i\frac{\sqrt 2}{2}$, it goes back to linearly polarized wave (\ref{LPTW2}).
It is thus clear that the rotation of the polarization plane of linearly polarized wave (\ref{LPTW1}) or (\ref{LPTW2}), when considered as a coherent superposition of RCP and LCP waves, comes from the rotation of the polarization states of their circularly polarized components.

\section{Conclusions and Remarks}\label{conclusions}

In conclusion, we showed through a detailed logical analysis that there is not circular birefringence in an isotropic chiral medium. We proved that both the RCP and the LCP waves propagate at the same phase velocity as the linearly polarized wave. In addition, we found that any elliptically polarized wave in a particular chiral medium is rotated in the same way.
This is described by its rotatory polarization vector and is indicated by expression (\ref{a-LPBs}) or, equivalently, by expression (\ref{a-CPBs}).
The circularly polarized waves are so special that the result of the rotation of their polarization states appears as opposite phases as if, as expressed by Eqs. (\ref{E+t_no}) and (\ref{E-t_no}), they propagated at different phase velocities with their polarization states transmitted unchanged.

What is remarkable is that the rotation of the polarization state of the plane wave (\ref{Et-LPBs}) in the chiral medium comes only from the rotatory polarization bases (\ref{uplusv}). It does not involve the change of the Jones vector. 
This is in consistency with the observation \cite{Barr, Alex-BLVY, Xi-WWB, Chen-R} that the optical rotation in the chiral medium shows up as the rotation of the axes of polarization ellipse. It is hence improper to describe the optical activity as the transformation of the Jones vector \cite{Dama}.
The optical activity in the chiral medium implies a new mechanism to change the state of polarization of light. It amounts to the change of the polarization bases with the Jones vector remaining fixed.
The physics that underlies such a mechanism needs further discussions, which are beyond the scope of the present paper and will be presented elsewhere.

\section*{Acknowledgments}

This work was supported in part by National Natural Science Foundation of China under Grant No. 11974251.


\begin{thebibliography}{99}
\bibitem{Note}
Here circular double refraction refers to the phenomenon that the refracted wave consists of RCP and LCP components propagating in different directions. It does not mean circular birefringence, which refers to different velocities or different indices of refraction for RCP and LCP components.
\end{thebibliography}
\end{document}